\documentstyle[12pt,aasms4]{article}
\begin{document}

\title{Detection of a new methanol maser line with the Kitt Peak 12--m
telescope by remote observing from Moscow.}

\author{V.I.~Slysh$^1$\altaffiltext{1}{Astro Space Center, Lebedev 
Physical Institute, Profsoyuznaya str. 84/32, 117810, Moscow, Russia;
vslysh@dpc.asc.rssi.ru; kalensky@socrat.asc.rssi.ru; 
ivaltts@dpc.asc.rssi.ru.}, S.V.~Kalenskii$^1$, I.E.~Val'tts$^1$
and V.V.~Golubev$^2$\altaffiltext{2}{Astronomical Division, Moscow State 
University, Vorob'evy Gory, 119899, Moscow, Russia; golubev@dpc.asc.rssi.ru.}}

\begin{abstract}
A new methanol maser line $6_{-1}-5_0~E$ at 133~GHz was detected with the 
12--m Kitt Peak radio telescope using remote observation mode from
Moscow. Moderately strong, narrow maser lines were found in DR21(OH), 
DR21--W, OMC--2, M8E, NGC2264, L379, W33--Met. The masers have similar
spectral features in other transitions of methanol$-E$ at
36 and 84~GHz, and in transitions of methanol$-A$ at 44 and 95~GHz.
All these are Class~I transitions, and the new masers also
belong to Class~I. In two other methanol transitions near 133~GHz,
$5_{-2}-6_{-1}~E$ and $6_2-7_1~A^+$, only thermal emission was detected
in some sources. Several other sources with wider lines in the
transition $6_{-1}-5_0~E$ also may be masers, since they do not
show any emission at the two other methanol transitons near 133~GHz.
These are NGC2071, S231, S255, GGD27, also known as Class~I masers.
The ratio of intensities and line widths of the 133~GHz masers and
44~GHz masers is consistent with the saturated
maser model, in which the line rebroadening with respect to
unsaturated masers is suppressed by cross relaxation due
to elastic collisions.
\end{abstract}
   \section{Introduction}
The widespread galactic methanol masers were found in many
transitions in the microwave band. Two kinds of sites are associated
with methanol masers. Cold dust cores sometimes containing
star formation regions at a very early, pre--stellar stage of
evolution are associated with Class I methanol masers. They emit in a
particular set of methanol transitions at 25, 36, 44, 84, and 95~GHz.
Another kind of star formation regions with young stars which are
embedded in nascent molecular cores and produce ultra--compact HII
regions, is associated with methanol masers of Class~II. They emit lines
in a different from Class~I set of transitions, typically, at
frequencies 6.7 and 12~GHz. Recently a new Class~II methanol maser line
was detected at a very high frequency 157~GHz (Slysh et al. 1995) (see Menten (1991) for an earlier discussion of methanol maser classification). 
   Masers of Class~I in methanol--$E$ were found in transitions
   between the levels of the $K=-1$ stack and those of the $K=0$ stack.
   Zuckerman et al. (1972) proposed a
   model of maser inversion by collisional excitation of molecules,
   followed by radiative decay. A much faster spontaneous
   decay of the $K=0$ levels leads to their underpopulation relative to 
   the $K=-1$ levels, which results in the inversion of the transitions 
   between the $K=-1$ and $K=0$ levels. Masers are expected in transitions 
   $J_{-1}-(J-1)_0~E$ with $J\ge 4$, when the $J_{-1}$ 
   levels become higher than the $(J-1)_0$ levels. The first transition of
   this type is $4_{-1}-3_0~E$, at 36 GHz, and was discovered as a 
   maser by Morimoto et al. (1985) in Sgr B2. The second transition, 
   $5_{-1}-4_0~E$, was discovered at 84~GHz as a strong maser by 
   Batrla and Menten (1988) in DR21(OH). The next transition 
   is $6_{-1}-5_0~E$, at 133~GHz. The methanol line in this transition 
   was first observed by Cummins et al. (1986) from Sgr B2, but it was
   not clear if it was a maser line. In this paper, we report on the 
   detection of the maser emission from many sources in the transition
   $6_{-1}-5_0~E$ at 133~GHz.   
   \section{Observations}
   The observations were made on June 5~--- 7, 1995 with the 
   12--m NRAO\footnote[1]{NRAO is operated by Associated Universities, Inc., 
   under contract with the National Science Foundation.} 
   telescope at Kitt Peak. For the first time, the 
   observations were conducted in remote observing mode from the Astro 
   Space Center in Moscow. For this purpose, a Sun workstation at the Astro 
   Space Center was connected via Internet to the computer of the 12--m
   telescope. An astronomer in Moscow could see on the screen of his
   computer the status of the telescope, the weather information, and a 
   TV monitor of the telescope inside the dome and environs, and was 
   also able to converse with the telescope operator using the 'talk'
   procedure. The spectra 
   were monitored in real time using the UNIPOPS software package. The
   communication between the telescope and Astro Space Center was 
   rather stable, with only one short break during the 48--hour 
   observing session. The observations of the methanol line were 
   carried out at 
   the rest frequency of the $6_{-1}-5_0~E$ transition, 132890.790~MHz
   (Lees et al. 1973), with the dual--polarization SIS receiver.
   Two other methanol lines were also observed near this frequency:
   $5_{-2}-6_{-1}~E$ at 133606.5~MHz and $6_2-7_1~A^-$ at 132621.94~MHz.
   The typical
   system noise temperature was 200 to 250~K. The observations were 
   performed in position--switching mode. The beamwidth was $41''$,
   and the pointing accuracy was better than $5''$. The calibration 
   was achieved by the standard vane method, and for a point source, 1~K of 
   $T_R^*$ was equal to 35 Jy. A hybrid spectrometer with 37.5~MHz
   total bandwidth and 384 channels in each polarization was used, 
   providing velocity resolution of 0.11~km~s$^{-1}$.
   \section{Results}
   Forty--one sources were observed with a detection 
   limit of 1.75~Jy. The list of sources included known methanol masers both
   of Class~I and Class~II. The emission line of methanol was detected
   from 35 sources, the non--detections being Class~II methanol masers
   associated with IRAS sources. In the rest of the Class~II methanol masers,
   thermal emission was detected. On the contrary, many of the Class~I
   masers showed intense narrow, evidently maser lines at 133~GHz. 
   We tentatively define as masers sources which show narrow
   lines with line width less than 1~km~s$^{-1}$.
   An additional evidence of the maser emission would be its high
   brightness temperature, exceeding, e.g. 10$^4$~K. Unfortunately, present
   observations were made with a low angular resolution provided by the
   41$''$ beam of the 12--m Kitt Peak telescope. If one assumes that the
   sources were unresolved by the telescope beam with the upper limit of
   the angular size of 20$''$, then the lower limit of the brightness
   temperature even for the stronger sources DR21(OH) and DR21--W is only
   150~K. Observations with a much higher angular resolution at millimeter
   interferometers such as the BIMA array or Plateau de Bure interferometer
   are needed in order to firmly establish the maser nature of the emission
   lines. Here we report 
   on new maser sources defined as above. Their spectra are shown in Fig.1, 
   and the Gaussian line parameters are given in Table~1.
   Results on the rest of
   the sources will be reported elswhere. In the transitions $5_{-2}-6_{-1}~E$
   and $6_2-7_1~A^-$, predicted by Cragg et~al. (1992) as Class~II methanol
   masers, no lines were detected in the new maser sources
   except for weak wide lines in DR21(OH).
   \subsection{Notes on individual sources}
   \hfil\break $\bf{OMC-2}$. The $6_{-1}-5_0~E$ spectrum (Fig.1) can be 
   approximated 
   by a strong, narrow, most probably maser line, and a weak broad pedestal 
   line. The narrow line is at the same radial velocity 
   as maser lines in the $7_0-6_1~A^+$ transition at 44~GHz 
   and the $4_{-1}-3_0~E$
   transition at 36~GHz (Haschick, Menten, and Baan 1990), the  
   $5_{-1}-4_0~E$ transition at 84.5~GHz (Menten 1991), the $5_2-5_1~E$
   and $6_2-6_1~E$ transitions at 25~GHz (Menten et al. 1988b), and the
   $8_0-7_1~A^+$ transition at 95~GHz (Menten 1991, Val'tts et al. 1995). 
   The line width 
   is somewhat larger (0.68~km~s$^{-1}$) at this frequency than at 44~GHz
   (0.35~km~s$^{-1}$), where flux density is the highest. The broad 
   component was not seen in these transitions but was observed by
   Menten et al. (1988a) in $2_k-1_k$ thermal lines of methanol at 96~GHz. 
   \hfil\break   $\bf{NGC2264}$. The $6_{-1}-5_0~E$ spectrum (Fig.1)
   also shows a superposition of a narrow maser line and a broad 
   line at close radial velocities. A similar two--component
   spectrum was observed in the $4_{-1}-3_0~E$ transition at 36~GHz (Haschick,
   Menten, and Baan 1990), the $5_{-1}-4_0~E$ transition at 84.5~GHz (Menten 
   1991), and
   the $8_0-7_1~A^+$ transition at 95~GHz (Val'tts et al. 1995).
   In the transition $7_0-6_1~A^+$ at 44~GHz there are two
   narrow lines separated by 0.28~km~s$^{-1}$ (Haschick, Menten, and Baan 
   1990). At 132~GHz the line is broader, 0.84~km~s$^{-1}$, and may
   be a blend of two narrow lines.
   See also the collection of spectra at frequencies from 36 to 146~GHz of
   Menten (1991). Menten also shows a broad absorption profile at 12.1~GHz
   ($2_0-3_{-1}~E$) with width and shape similar to the broad
   emission at 36~GHz.
   \hfil\break   $\bf{M8E}$. This methanol maser was first detected at 
   36~GHz (Kalenskii et al. 1994), then found to be one of the most
   powerful masers at 44~GHz (Slysh et al. 1994). At 133~GHz, it is not 
   so strong, with a flux density of about 60~Jy; the 
   profile is something broader and consists of an intense narrow  
   component and a weak, broad component (Fig.1).
   \hfil\break   $\bf{W33-Met}$. This source has a narrow maser line at a
   radial velocity of 32.8~km~s$^{-1}$ and a broad line shifted
   by 4~km~s$^{-1}$ to higher velocities (Fig.1). At other methanol transition
   frequencies, there is a narrow maser line at 9.9~GHz 
   in the transition $9_{-2}-8_{-1}~E$ (Slysh et al. 1992);
   at 25~GHz, the maser lines are seen at the radial velocity 33.2~km~s$^{-1}$
   in the transitions $J_2-J_1~E$ with $J$=3 to~6, and 9, 
   at the radial velocity 36.3~km~s$^{-1}$,
   there is a broad line in transitions $J$=2 to~6 
   (Menten et al. 1986). At 36~GHz ($4_{-1}-3_0~E$), the spectrum is similar,
   with the narrow maser line and broad line red shifted by
   4.2~km~s$^{-1}$ (Kalenskii et al. 1994). Similar two--component 
   spectra were observed at 44~GHz ($7_0-6_1~A^+$) by Haschick, Menten
   and Baan (1990) and at 95~GHz ($8_0-7_1~A^+$) by Val'tts et al. (1995).
   Pratap and Menten (1993) mapped the 95~GHz maser and found that its
   position coincides with the position of the 25~GHz maser to within
   about 2$''$.
   \hfil\break   $\bf{L379.}$ The $6_{-1}-5_0~E$ spectrum (Fig.2, left 
   middle spectrum) shows a double--peaked line with intense wings.
   The line has been separated into four Gaussian components, with 
   two narrow, probably maser lines, with 0.8 and 1.2~km~s$^{-1}$
   line width (Table~1). The methanol maser in L379 was first
   discovered at 44~GHz and 36~GHz (Kalenskii et al. 1992). The shape
   of the line at 36~GHz with two peaks is similar to the shape at 
   133~GHz. A moderately
   strong line was found at 95~GHz (Val'tts et al. 1995) with line
   width 3.9~km~s$^{-1}$. This rather large line width probably
   results from a blend of several narrow maser lines.
   Other evidence of the maser nature of the
   133~GHz emission comes from a comparison of the intensity of this 
   line with the two other methanol lines observed in this experiment.
   In Fig.2 (left) spectra of the three methanol transitions near 
   133~GHz are presented in comparison with a similar set of spectra
   of Ori--KL (Fig.2, right). While in Ori--KL the lines are present
   in all three transitions with relative intensities proportional
   to the line strength that is typical for the thermal emission,
   in L379 only the $6_{-1}-5_0~E$ line is present,
   and the two other lines are weaker than 0.03 of this line. This is
   consistent with a non--thermal distribution of line intensities
   and maser amplification in the $6_{-1}-5_0~E$ line.
   Several sources also show only the $6_{-1}-5_0~E$ line with a 
   moderately narrow line width, and are probably relatively weak
   masers: NGC2071, S255, S231, GGD27. They are Class~I methanol
   masers in other transitions.   
   \hfil\break   $\bf{DR21(OH)~and~DR21-W.}$  These two strongest
   133~GHz masers separated by 3$\lefteqn{'}$.6 show very intense 
   narrow maser lines superposed on a broad component (Fig.1).
   The narrow maser
   line in DR21(OH) was observed in other methanol transitions at
   36, 44, 84, 95~GHz, and in DR21--W at 25, 36, 44, 84, 95~GHz  
   (Menten et al. 1986, Haschick et al. 1990, Batrla and Menten 1988,
   Plambeck and Menten 1990, Menten 1991) at the same radial velocity. 
   The broad
   line was observed in DR21(OH) at 25, 36, 44, 84 and 95~GHz, and
   in DR21--W only at 95~GHz. In DR21(OH) broad lines in $5_{-2}-6_{-1}~E$
   and $6_2-7_1~A^-$ transitions were found at the same radial velocity
   as the broad component in $6_{-1}-5_0~E$ transition, with the intensity
   of about 5 per cent of the $6_{-1}-5_0~E$ line intensity.
   \section{Discussion}
   The similarity between line profiles of the new 133~GHz methanol
   masers and methanol masers of Class~I as well as association
   with the same sources, suggests that the 133~GHz methanol masers
   belong to the Class~I. The corresponding methanol transition $6_{-1}-5_0~E$
   is of the same type $K=-1$ $\to$ $K=0$, as most of the Class~I
   transitions in methanol--$E$.
   The detection of new masers in the transition $6_{-1}-5_0~E$
   supports the suggestion that $K=-1$ ladder levels are overpopulated
   relative to $K=0$ ladder levels, which is the cause for the inversion
   of $J_{-1}-(J-1)_0$ transitions. A natural way of overpopulating
   the $K=-1$ ladder is collisional excitation followed by 
   spontaneous decay. The decay rates are different for the two
   ladders: for the upper $6_{-1}$ level the spontaneous decay rate is
   1.07$\times$10$^{-4}$~s$^{-1}$, while for the lower $5_0$ level it is 
   a factor of two higher 
   2.12$\times$10$^{-4}$~s$^{-1}$. The molecules excited by collisions to the
   $6_{-1}$ level remain there longer than at the $5_0$ level, resulting in 
   the higher population of the $6_{-1}$ level. The higher population of the 
   upper level in the $6_{-1}-5_0~E$ transition, or population inversion
   (relative to the normally lower population of the  upper level in
   thermally excited molecules) may produce the maser emission if the
   column density is high enough to provide significant amplification. The 
   same mechanism of level population inversion holds for other similar 
   transitions in methanol--$E$: $4_{-1}-3_0$, $5_{-1}-4_0$, $7_{-1}-6_0$,
   $8_{-1}-7_0$, and so on. In methanol--A, the $7_0-6_1$, $8_0-7_1$, 
   $9_0-8_1$ etc. transitions can be inverted in the same manner.
   A comparison of 133~GHz masers with the strongest Class~I methanol
   masers at 44~GHz shows that both show emission features at the same radial
   velocities. One can suppose that the emission features at both frequencies
   originate from the same region. Compared to the 44~GHz masers, the 133~GHz
   masers are weaker: the flux density is a factor 
   of 5 to 27 lower, and the line width is a factor of 1.4 to 4 larger.
   The flux density ratio is roughly consistent with the calculated optical
   depth ratio of about 6; this is expected for saturated masers at both
   frequencies. On the other hand, the observed larger line width at 133~GHz
   as compared to 44~GHz is more 
   appropriate for unsaturated masers, where the line width is inversely
   proportional to the square root of the optical depth. This apparent
   contradiction can be resolved if the masers are saturated but the line
   rebroadening due to the saturation does not take place, because the
   velocity distribution undergoes cross relaxation toward a Maxwellian 
   distribution
   as a result of elastic collisions (Nedoluha and Watson 1988). Therefore,
   line narrowing can be expected in the saturated masers, also; in the 
   case of Class~I methanol masers, collisions are important both for the 
   maser pump and for the velocity distribution relaxation, which permits 
   line narrowing.
   \section{Summary}
   The first successful remote observations from Moscow with the NRAO Kitt Peak
   radio telescope have resulted in the detection of new Class~I methanol masers 
   at 133~GHz. The millimeter--wave masers are as widespread as the 
   longer--wavelength masers: 35 sources were detected in emission; at least 
   seven are definite masers, and several more are good candidates. Two other
   methanol lines were used to discriminate between maser and thermal emission.
   The properties of the new maser sources are better described by the saturated 
   maser model, although the line width arguments require cross relaxation to
   explain the lack of line rebroadening. 
   \acknowledgments
   We are grateful to Dr.~Phil Jewell, Dr.~K.~Mead and the
   staff of the NRAO Kitt Peak radio telescope for help with the
   observations.
   The Sun Sparc station used in the remote observations
   was made available to the Astro Space Center through a grant from the 
   National Science Foundation to the Haystack Observatory.
   This work was supported by the Russian Foundation for Basic 
   Research (grant 95--02--05826).
   \newpage
   \figcaption{Spectra of the newly detected 133~GHz methanol--$E$
   masers.}
   \figcaption{133~GHz methanol spectra in three transitions for 
   L379~(left)
   and Ori~KL~(right). Note the anomalously large relative strength of the 
   $6_{-1}-5_0~E$ maser line in L379 as compared to the apparently
   thermal line ratio in Ori~KL.}
   
   \newpage
   \hfil\break Table~1. Line Parameters Determined from $6_{-1}-5_0~E$
Observations
\begin{tabular}{lrrrrr}
\tableline
~Name&~~R.A.(1950)&~~~Dec.(1950)&S(Jy)~~~&V$_{lsr}$(km~s$^{-1}$)&
~~$\Delta$V(km~s$^{-1}$) \nl
\tableline
OMC--2&05 32 59.8&$-$05 11 29&36.8(1.0)&11.39(0.01)&0.68(0.02)\nl
     &\dots &\dots &7.8(0.1)&11.66(0.07)&2.50(0.19)\nl
NGC2264&06 38 24.9&09 32 28&11.3(0.8)&7.54(0.02)&0.84(0.07)\nl
     &\dots &\dots &17.4(0.6)&7.94(0.04)&3.77(0.10)\nl
M8E&18 01 49.7&$-$24 26 56&12.9(1.2)&10.61(0.06)&2.68(0.16)\nl
     &\dots &\dots &55.2(1.3)&10.92(0.01)&0.77(0.02)\nl
W33--Met&18 11 15.7&$-$17 56 53&15.0(0.6)&32.78(0.02)&1.02(0.05)\nl
     &\dots &\dots &23.3(0.3)&36.76(0.02)&3.57(0.06)\nl
L379&18 26 32.9&$-$15 17 51&11.6(0.5)&16.39(0.15)&12.63(0.26)\nl
     &\dots &\dots &10.5(0.6)&17.69(0.02)&0.80(0.06)\nl
     &\dots &\dots &40.9(0.7)&18.96(0.03)&4.81(0.06)\nl
     &\dots &\dots &18.8(0.6)&20.24(0.02)&1.20(0.05)\nl
DR21--W&20 37 07.6&42 08 46&11.3(0.5)&$-$2.76(0.06)&4.20(0.15)\nl
     &\dots &\dots &64.7(0.8)&$-$2.47(0.01)&0.62(0.01)\nl
DR21(OH)&20 37 13.8&42 12 13&63.4(1.6)&0.28(0.01)&0.86(0.02)\nl
     &\dots &\dots &56.7(0.6)&$-$2.34(0.04)&5.65(0.08) \nl
\tableline
\end{tabular}

\hfill\break Notes: Errors are 1$\; \sigma$ deviations determined 
from Gaussian fits.


\begin{references}
   \reference {a} Batrla,~W., \& Menten,~K.M., 1988, ApJ, 329, L117 
   \reference {b} Cragg,~D.M., Johns,~K.P., Godfrey,~P.D., \& Brown,~R.D., 1992,
   MNRAS, 259, 203
   \reference {c} Cummins, Sally,~E., Linke,~R.A., \& Thaddeus,~P., 
   1986, ApJS, 60, 819
   \reference {d} Haschick,~A.D., \& Baan,~W.A., 1989, ApJ, 339, 949
   \reference {e} Haschick,~A.D., Menten,~K.M., \& Baan,~W.A., 1990, ApJ, 354, 556
   \reference {f} Kalenskii,~S.V., Bachiller,~R., Berulis,~I.I., Val'tts,~I.E.,
   Gomez--Gonzalez,~J., Martin--Pintado,~J., Rodriguez--Franco,~A., \& Slysh,~V.I.,
   1992, Astron.Zh., 69, 51
   \reference {g} Kalenskii,~S.V., Berulis,~I.J., Val'tts,~I.E., Dzura,~A.M., 
   Slysh,~V.I., \& Vasil'kov,~V.I., 1994, Astron.Zh., 71, 51
   \reference {h} Lees,~R.M., Lovas,~F.J., Kirchhoff,~W.H., \& Johnson,~D.R., 
   1973, J.~Chem.~Ref.~Data, 2, 205
   \reference {i} Menten,~K.M., 1991, in "Skylines", Proceedings of the Third
   Haystack Observatory Meeting, ed. Haschick,~A.D., \& Ho,~P.T.P., San
   Francisco: Astronomical Society of the Pacific
   \reference {j} Menten,~K.M., Walmsley,~C.M., Henkel,~C., \& Wilson,~T.L.,
   1986, A\&A, 157, 318
   \reference {k} Menten,~K.M., Walmsley,~C.M., Henkel,~C., \& Wilson,~T.L.,
   1988a, A\&A, 198, 253
   \reference {l} Menten,~K.M., Walmsley,~C.M., Henkel,~C., \& Wilson,~T.L.,
   1988b, A\&A, 198, 267
   \reference {m} Morimoto,~M., Ohishi,~M., \& Kanzawa,~T., 1985, ApJ, 288, L11
   \reference {n} Nedoluha,~G.E., \& Watson,~W.D., 1988, ApJ, 335, L19
   \reference {o} Plambeck,~R.L., \& Menten,~K.M., 1990, ApJ, 364, 555
   \reference {p} Pratap,~P., \& Menten,~K., 1993, in "Astrophysical Masers", eds
   Clegg,~A.W., \& Nedoluha,~G.E., Lecture Notes in Physics, Springer--Verlag
   \reference {q} Slysh,~V.I., Kalenskii,~S.V., \& Val'tts,~I.E., 1992, ApJ, 397, L43
   \reference {r} Slysh,~V.I., Kalenskii,~S.V., \& Val'tts,~I.E., 1994, MNRAS, 
   268, 464
   \reference {s} Slysh,~V.I., Kalenskii,~S.V., \& Val'tts,~I.E., 1995, ApJ, 442, 668
   \reference {t} Val'tts,~I.E., Dzura,~A.M., Kalenskii,~S.V., Slysh,~V.I.,
   Booth,~R., \& Winnberg,~A., 1995, A\&A, 294, 825   
   \reference {u} Zuckerman,~B., Turner,~B.E., Johnson,~D.R., Palmer,~P., \&
   Morris,~M., 1972, ApJ, 177, 601
   \end{references}
\end{document}